\documentclass[aps,pra,showpacs,twocolumn]{revtex4-1}
\usepackage{graphicx}
\usepackage[usenames]{color}
\usepackage{amssymb}

\newcommand{\fig}[1]{Fig.~\ref{#1}}
\newcommand{\be}[1]{\begin{equation}\label{#1}}
\newcommand{\ee}{\end{equation}}
\begin{document}

\title{Prevalence of different double ionization pathways and traces of three-body collisions in strongly driven Helium}

\author{A. Emmanouilidou}
\address{Department of Physics and Astronomy, University College London, Gower Street, London WC1E 6BT, United Kingdom.\\
and Chemistry Department, University of Massachusetts at Amherst, Amherst, Massachusetts, 01003, U.S.A.
}

\begin{abstract}
We present a unified treatment of the prevalence of different double ionization (DI) pathways as a function of the sum of the final electron energies in strongly driven Helium. We do so as a function of laser frequency and intensity. At small total energy a new DI pathway prevails where both electrons are ionized with a delay after re-collision.  We find that three-body collisions between the two electrons and the nucleus underly this pathway--we refer to it as triple collision (TC) pathway. At high total energy the pathway that prevails is the one where one electron is ionized with a delay following re-collision---we refer to it as Delayed pathway. Asymptotic observables that trace the prevalence of the latter pathways are identified. 
\end{abstract}
\date{\today}
\maketitle

\section{Introduction}
The Helium atom driven by strong infrared laser fields has attracted considerable interest over the last two decades. The competing forces in strongly driven He, namely, the influence of the laser field on each electron,  the electron-electron repulsion, and the interaction of each electron with the nucleus result in a wealth of physical phenomena---many still unexplored. 
When the field intensity is large the two electrons are stripped out sequentially \cite{lambropoulos1985} while for smaller intensities non-sequential double ionization (NSDI) dominates. In this work, we address the latter intensity regime.

According to the three-step, also called re-scattering, model \cite{Corkum93PRL}---the accepted mechanism for NSDI---1) one electron escapes through the field-lowered Coulomb potential, 2) it moves in the strong infrared laser field and 3) it returns to the core to transfer energy to the other electron remaining in He$^{+}$. Theoretical support for the three step model has been provided, among others, by \cite{ruiz2006, haan2008, ParkerJPB2003,Prauzner}.
Although the re-scattering model has worked well in providing the interpretation of the basic strong field phenomena, such as ATI (Above Threshold Ionization), HHG (High-Order Harmonic Generation) and NSDI (Non-Sequential Double Ionization), recent kinematically complete
experimental investigations \cite{staudte2007,rudenko2007}, have revealed additional structure in the latter for 800 nm laser fields. Specifically they observed the so-called ``finger-like'' structure (V-shape) in the correlated momenta of the outgoing electrons, previously predicted  by quantum mechanical studies of He DI at 400 nm \cite{parker2006}. Recent classical \cite{EmmanouilidouPRA2008a, fingerlike} and quantum-mechanical \cite{CDLinPRL2010} theoretical treatments have added to our understanding of the latter.

A still open problem is a unified description of the different DI pathways the two electrons follow to escape after re-collision. The Direct and Delayed are two well established pathways: 
one involves an almost simultaneous ionization (SI) of the two electrons and the other an ejection with a delay of approximately one quarter of a laser period or more (The delayed pathway is also referred to as re-collision-induced excitation with subsequent field ionization, RESI \cite{KopoldPRL2000,FeuersteinPRL2001,Carla}).
 Taking a first step towards a general description of the DI pathways, in ref.  \cite{EmmanouilidouNJP} we used full-dimensional classical and quantum mechanical techniques for He at 400 nm. After establishing a remarkable agreement of the two treatments we used the classical calculation to explore the features of the DI mechanisms in three different regimes of the total electron energy. 
We found that the delayed pathway prevails for small intensities---not addressed in the current work---independently of total electron energy. (Throughout this work the total electron energy we refer to is the sum of the final kinetic energies of the two electrons).  In contrast, at higher intensities the SI pathway prevails
up to an upper-limit in total energy that shifts upwards with increasing intensity. The effect of the nucleus proved crucial for explaining this upward shift.

In this work, in a quasiclassical framework, we explore the prevalence of the DI pathways as a function of total energy in energy steps as small as 
 the immense computational challenge of the endeavor allows---600-700 thousand of DI events for the whole energy regime.
We do so as a function of laser intensity and for two frequencies corresponding to wavelengths of 400 nm and 800 nm. We also identify asymptotic observables where the prevalence of a DI pathway over the rest can be clearly traced.
This general treatment on one hand allows us to explore whether previous findings at 400 nm are valid for 800 nm and 
on the other reveals new ones. 

One such new finding is that at very small total energy a new DI pathway prevails. We refer to it as TC (triple collision) with both electrons ionizing more than a quarter of a laser cycle after re-collision. TC is present for all laser frequencies and intensities considered in the current study. We find that {\it three-body collisions between the two electrons and the nucleus underly the TC pathway}. 
In addition, at high total energy we find that the Delayed pathway overtakes the Direct one during a very sharp transition. These findings hold true for a range of different intensities and frequencies, namely, for 9/13$\times$10$^{14}$Watts/cm$^2$ at 400 nm and for 2.25/3.25$\times$10$^{14}$Watts/cm$^2$ at 800 nm. 

Besides the above mentioned features common to DI at 400 nm and 800 nm, we identify a frequency dependent feature. The number of times the re-colliding electron returns to the core before both electrons are ionized decreases with increasing frequency. (This number is defined through electron returns roughly at times (2/3+n)T---three-step model, with T the period of the field). This difference
roughly accounts for the different patterns of the correlated momenta for the Delayed mechanism observed at 400 nm and 800 nm.

\section{Model} 
The three-dimensional quasiclassical model we use entails one electron
tunneling through the field-lowered-Coulomb potential
with a quantum tunneling rate given by the ADK formula \cite{ADK}. The
laser pulse used in the classical calculations is  $E(t) = E_{0}(t) \cos(\omega t)$ and is linearly
polarized along the z-axis.
The longitudinal momentum is zero while the
transverse one is given by a Gaussian distribution. The remaining electron is modeled by a microcanonical
distribution \cite{microcanonical}.
 The pulse envelope is defined as $E_{0}(t) = E_{0}$ (a constant) for $0 < t < 6 {\rm T}$
and $E_{0}(t) = E_{0} \cos^{2}(\omega (t-6{\rm T})/12)$ for $6{\rm T} < t < 9{\rm T}$ with T the period of the field.  
 An advantage of our classical propagation over other classical techniques is that we employ regularized coordinates \cite{regularized} (to
account for the Coulomb singularity) which results in a faster and more stable numerical propagation.

\section{Double Ionization mechanisms}
\subsection{\% Contribution of DI mechanisms as a function of total electron energy}
\begin{figure}[h]
{\includegraphics[scale=0.25,clip=true]{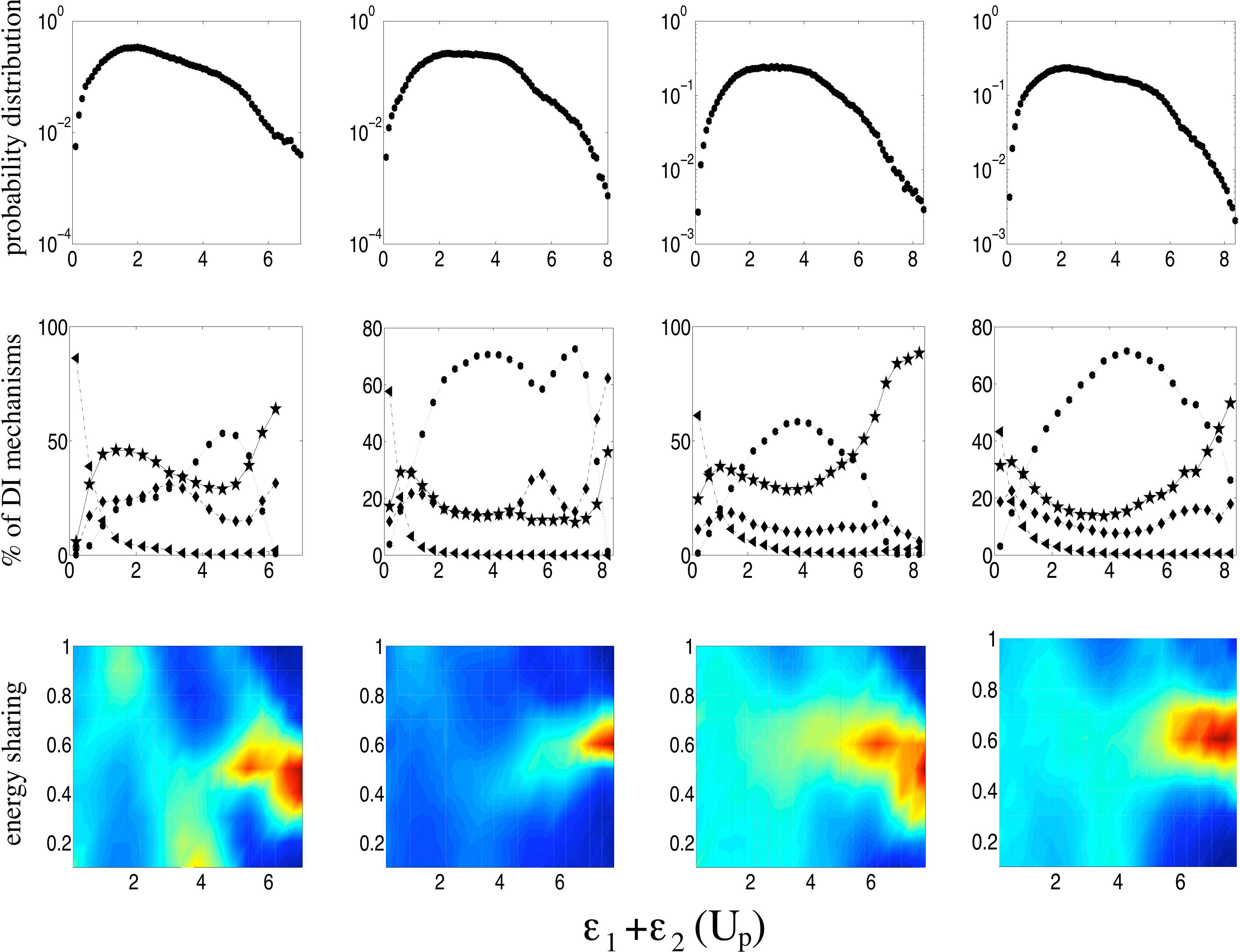}}
\caption{\label{fig:percent} First row: probability distribution for $\epsilon_{1}+\epsilon_{2}$; Middle row: \% of DI mechanisms with $\bullet$ for SI, $\blacklozenge$ for RESIa, $\star$ for RESIb and $\blacktriangleleft$ for TC; Bottom row: energy sharing, i.e., $\frac{|\epsilon_{1}-\epsilon_{2}|}{ |\epsilon_{1}+\epsilon_{2}|} $ as a function of  $\epsilon_{1}+\epsilon_{2}$. (From left to right) first column: 2.25$\times$10$^{14}$ Watts/cm$^2$ at 800 nm, second column: 3.25$\times$$10^{14}$ Watts/cm$^2$ at 800 nm, third column: 9$\times$$10^{14}$ Watts/cm$^2$ at 400 nm and fourth column: 13$\times$$10^{14}$ Watts/cm$^2$ at 400 nm.
  }
\end{figure}

In \fig{fig:percent} we show the percentage contribution of the DI mechanisms as a function of the sum of the two electron energies. We do so for laser pulse intensities of 2.25/3.25$\times$10$^{14}$Watts/cm$^2$ at 800 nm and for 9/13$\times$10$^{14}$Watts/cm$^2$ at 400 nm. The range of laser intensities we currently consider is important because, at 400 nm, for 9 (12) $\times 10^{14}$ W/cm$^2$  the maximum return energy of the re-colliding electron (3.2 $U_{p}$ 
according to the three-step model \cite{Corkum93PRL}) equals the first excitation (ionization) energy of ground state He$^{+}$ and similarly for 2.25 (3.25) $\times 10^{14}$ W/cm$^2$ at 800 nm. 
The ponderomotive energy $U_{p}=E_{0}^{2}/(4\omega^{2})$ is the average energy an electron gains in one laser cycle. 
The intensities 9 (13) $\times 10^{14}$ W/cm$^2$ at 400 nm correspond to the same ponderomotive energy as for 2.25 (3.25) $\times 10^{14}$ W/cm$^2$ at 800 nm, and thus the same Keldysh parameter $\gamma=\sqrt{I_{p}/(2U_{p})}$ \cite{Keldysh}. We adapt an energy step, for the final total energy, of 0.4 U$_{p}$ to achieve good statistics in all energy intervals.

To identify the main DI energy transfer pathways we use the time delay between the re-collision of the free electron with the parent ion and
the onset of ionization of the second electron \cite{EmmanouilidouPRA2009}. For a definition of the time of ionization see \cite{EmmanouilidouPRA2008a} and references there in. In the direct ionization
pathway (SI) both electrons are ionized simultaneously very close (less than a quarter of a laser period) to the re-collision time. In
the delayed ionization pathway, the re-colliding electron excites the remaining electron but does not ionize it.  The
electron is subsequently ionized at a peak (RESIa) or at a zero (RESIb) of the laser electric field \cite{KopoldPRL2000,FeuersteinPRL2001}. Finally in the TC pathway both electrons are ionized more than a quarter of a laser period following re-collision.  
The \% break-up of the different DI mechanisms as a function of the total electron energy exposes universal features present for different laser intensities and frequencies.

\subsection{A new pathway of DI ionization for small total electron energy}
   One such feature is that for small total energy the two electrons escape through the TC pathway; both electrons are ionized with a delay of more than a quarter of a laser cycle with respect to the re-collision time.  TC reaches already 86\% for 2.25$\times$10$^{14}$ Watts/cm$^2$ at 800 nm in the energy interval [0, 0.4] U$_{p}$, making this parameter
   regime ideal for the study that follows.

\begin{figure}[h]
 {\includegraphics[scale=0.4,clip=true]{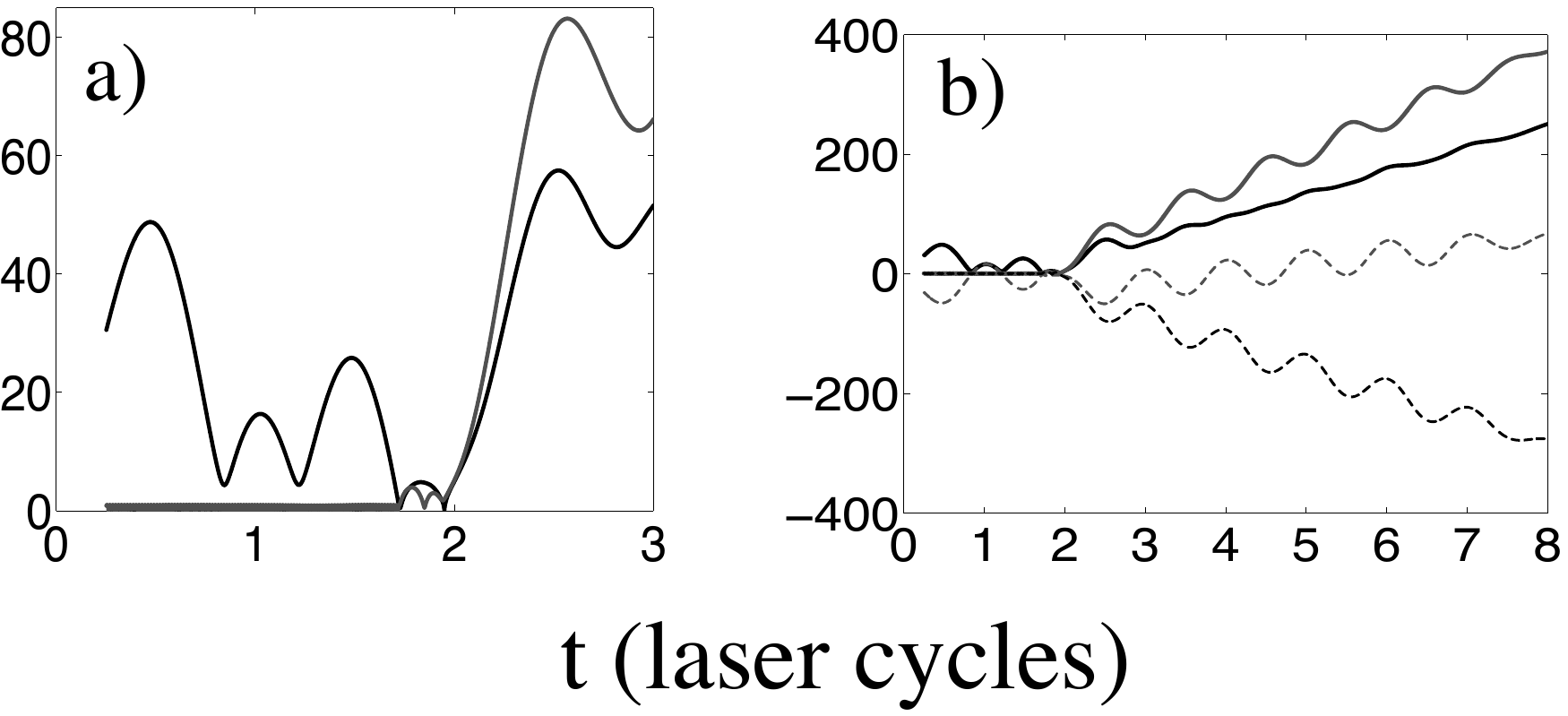}}
\caption{\label{fig:trajectory}  a) The radii of the re-colliding (black line) and of the initially bound electron (grey line); b) The distances in the direction of the laser polarization (z) for the re-colliding (black dashed line) and for the initially bound electron (grey dashed line) for TC for 2.25$\times$$10^{14}$ Watts/cm$^2$ at 800 nm. The re-collision time is around 1.75 laser cycles. Both electrons are ionized between 2-2.5 cycles.
  }
  \end{figure}

The signature feature of the TC pathway is {\it three-body collisions between the nucleus and the two electrons}. For the vast majority of  trajectories the re-colliding electron following re-collision gets temporarily ``trapped" orbiting the nucleus. In \fig{fig:trajectory} a) a representative TC trajectory is depicted where the first electron, following re-collision, orbits the nucleus once before both electrons are ionized. Trajectories where the first electron orbits the nucleus more than once are also common. The three body collision is best evidenced by the subset of trajectories where, in addition to orbiting the nucleus, the two electrons ionize at the same time, see \fig{fig:trajectory}; if the interaction of the laser field with the two electrons was the prevalent one then the electronic positions along the field polarization would be similar. Instead, the three-body collision prevails and since the energy shared among the three particles is small the two electrons escape in almost opposite directions along the field polarization. 

\begin{figure}[h]
 {\includegraphics[scale=0.3,clip=true]{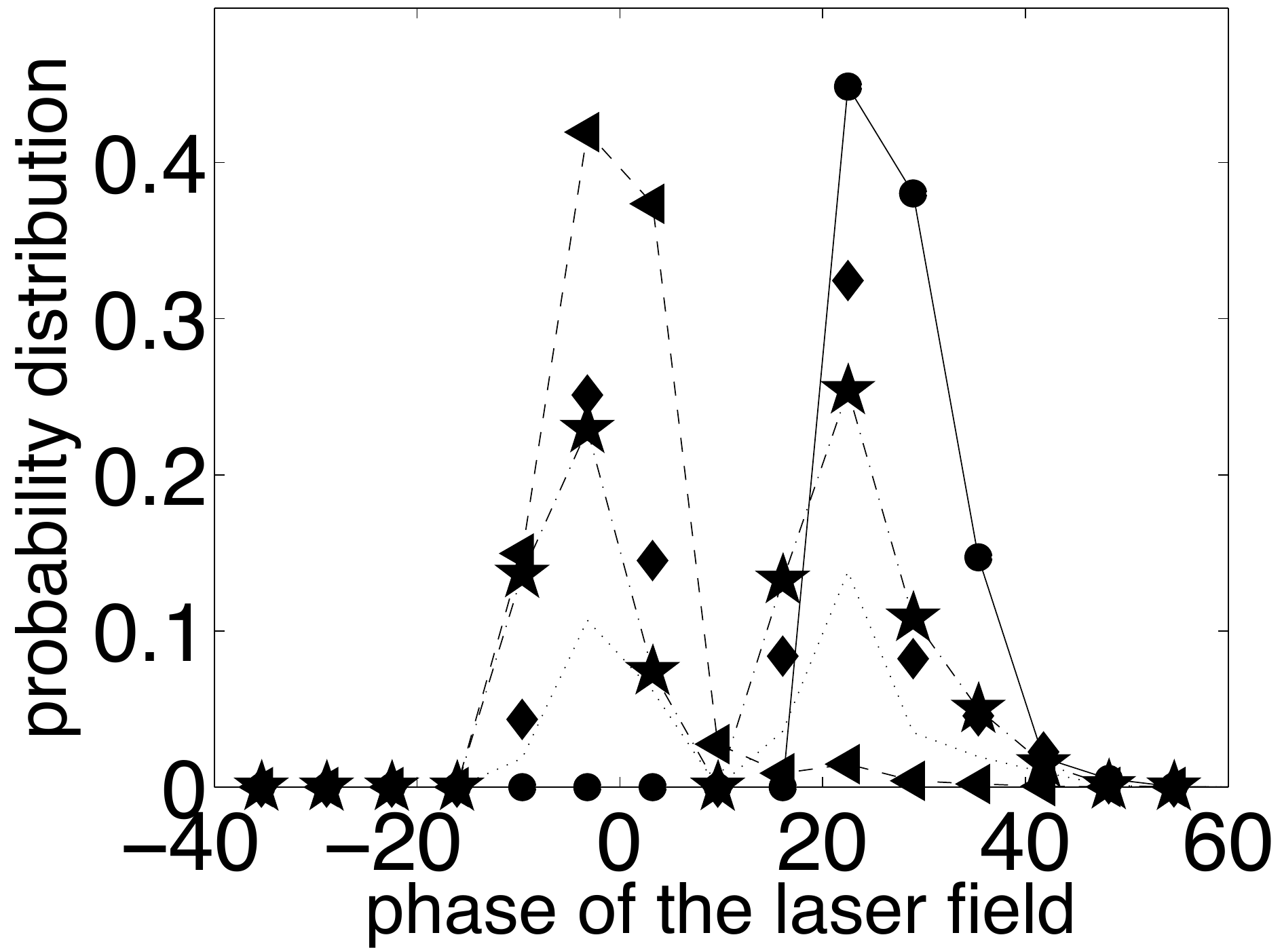}}
\caption{\label{fig:pow325}  Probability distribution for the re-colliding electron to tunnel at a given phase of the laser field; $\bullet$ for SI, $\blacklozenge$ for RESIa, $\star$ for RESIb and $\blacktriangleleft$ for TC for 800 nm and 2.25$\times$$10^{14}$ Watts/cm$^2$.
  }
  \end{figure}

To better understand why the first electron gets trapped after re-collision takes place, we identify the phase of the laser field when the first electron tunnels. In \fig{fig:pow325}, we show that in the TC pathway the re-colliding electron tunnels into the continuum when the phase of the laser field is around -3$^{\circ}$. This results in the re-colliding electron returning to the nucleus with small energy compared to the maximum return energy of 3.2 U$_{p}$ achieved when tunneling takes place around 17$^{\circ}$(three-step model). In contrast, in  the SI mechanism the re-colliding electron tunnels
 when the phase of the field is around 20$^{\circ}$, resulting in higher energy at the time of re-collision. 
The return to the nucleus with small energy seems to be a necessary condition for the signature feature of the TC pathway. We finally note that for the majority of the TC trajectories re-collision takes place half a cycle before the re-colliding electron ionizes and the second electron is ionized primarily at the same time as the first electron or half a cycle later.

\begin{figure}[h]
{\includegraphics[scale=0.25,clip=true]{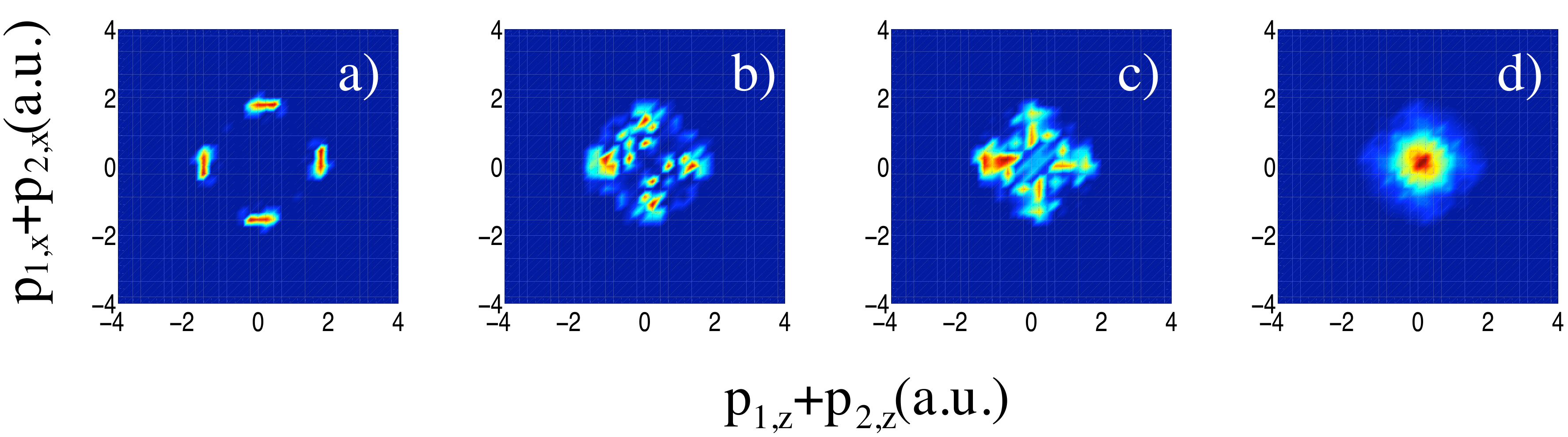}}
\caption{\label{fig:pow225}  DI probability as a function of the sum of the x component of the momenta versus the z component of the momenta, for 2.25$\times$$10^{14}$ Watts/cm$^2$ at 800 nm for the four different mechanisms SI (left), RESIa (second from left) 
RESIb (third from left) TC (right) for $\epsilon_{1}+\epsilon_{2}$ in the [0,0.4] $U_{p}$ energy interval. }
\end{figure}
\begin{figure}[h]
{\includegraphics[scale=0.25,clip=true]{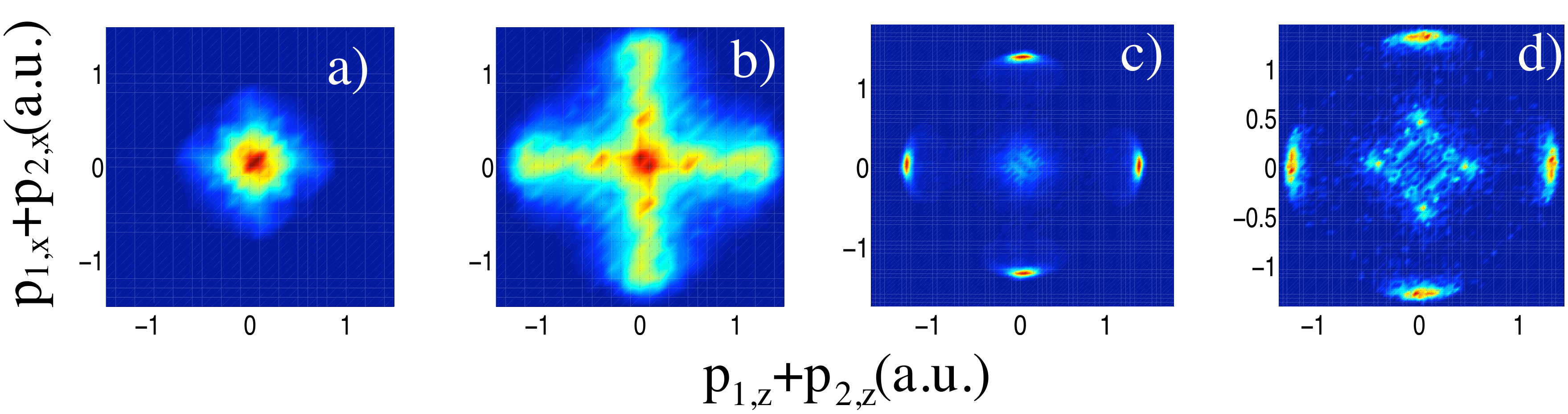}}
\caption{\label{fig:pow225all}  DI probability as a function of the sum of the x component of the momenta versus the z component of the momenta, for 2.25$\times$$10^{14}$ Watts/cm$^2$ at 800 nm for four energy intervals; from left to right
[0,0.4] $U_{p}$,  [0.8,1.2] $U_{p}$, [3.6, 4.0] $U_{p}$, [5.6, 6.0] $U_{p}$. The sum of the momenta is scaled by $1/\sqrt{E_{mid}}$, where $E_{mid}$ is the middle of the respective energy interval.}
\end{figure}

 The strong electron-electron repulsion is further evidenced using a triple differential distribution in \fig{fig:pow225}. For the smallest energy interval considered we compute the DI probability as a function of the sum of the momenta components parallel to the field polarization (z-axis) and perpendicular to it. From \fig{fig:pow225} d) we find that the highest DI probability for the TC pathway is when both the sum of the parallel and perpendicular components of the momenta are zero suggesting that a strong electron-electron repulsion underlies the two electron escape. This is not the case for the other pathways as shown in \fig{fig:pow225} a), b), c). 
Thus, one way to confirm experimentally the prevalence of the TC pathway for small energies would be to measure the triple differential probability described above 
in different energy regimes and find that only for the smaller energy interval the triple differential distribution resembles the TC one in \fig{fig:pow225} d). Indeed,  in \fig{fig:pow225all} we plot the correlated sum of the momenta in four energy regimes accounting for all doubly ionizing trajectories (all pathways included)
and find that only for the lowest energy interval, see \fig{fig:pow225all} a), the correlated sum of the momenta resembles that of \fig{fig:pow225} d). 

We have provided evidence that each electron is trapped by the nucleus after re-collision (\fig{fig:trajectory}), thus a strong electron-nucleus interaction takes place and we have shown that after re-collision there is a strong electron-electron repulsion (\fig{fig:pow225}, \fig{fig:pow225all}). We thus conjecture that triple collisions---interaction of the two electrons with the nucleus---underlies the pathway prevailing for small total electron energy. The TC pathway is a trace of collisional physics 
in strongly driven He, since a three-body interaction between two electrons and a nucleus dominates the threshold behavior of electron impact and photo-ionization processes, see Wannier's law \cite{Wannier}. 

Finally, let us note that the three-body collisions are more prevalent the smaller the interaction with the field is, explaining the decrease in the \% contribution of the TC pathway with increasing frequency and intensity, see \fig{fig:percent} second row.
 
\subsection{Direct versus Delayed pathways}
From \fig{fig:pow225all} we see that as the total energy increases the correlated sum of the momenta changes significantly indicating the prevalence of DI pathways other than the TC one. This is the focus of the current section. 

From \fig{fig:percent} we see that as the energy increases the Delayed pathway prevails up to total energy that differs with laser frequency and intensity; for larger intensities it extends up to around 1U$_{p}$ while for 2.25$\times$10$^{14}$ Watts/cm$^2$ at 800 nm up to 3U$_{p}$. The SI mechanism prevails for intermediate total energies
 extending up to approximately 5 $U_{p}$ for small intensities and up to around 7-8 $U_{p}$  
for higher ones, see \fig{fig:percent} and \cite{EmmanouilidouNJP}. This increase with intensity in the SI upper energy limit was attributed to more effective encounters of the re-colliding electron with the nucleus \cite{EmmanouilidouNJP}. The prevalence of the SI pathway is more pronounced for higher intensities reaching almost 70\%, see \fig{fig:percent}.  Increasing further the total energy we find that, through a sharp transition, the Delayed pathway prevails over the Direct one. The highest total energy we consider is 8 U$_{p}$ in order to ensure good statistics in all energy intervals.

The effect of the nucleus and the time of re-collision can be extracted from the classical trajectories by plotting for each DI pathway the average value of each electron's potential energy $-2/|\vec{r}_{1/2}|$ and momenta components parallel to the field polarization. As for 400 nm \cite{EmmanouilidouNJP}, we find that the re-collision time at 800 nm shifts from 0.5 T (T is the laser period) at small total energies to (n+2/3)T (three-step model), $n=0,1,2,3, ...$  at high total energies. The effect of the nucleus at 800 nm as for 400 nm is very important for total energies above approximately 5 U$_{p}$.   
However,  the effect of the nucleus is  also crucial for very small total energy as well where the TC pathway prevails, as we have shown above.

Another trace of collisional physics on strong field ionization is the overall shape of the SI contribution to DI as a function of total energy. The latter is very small for small and high energies while it is large for intermediate ones. Thus, SI qualitatively resembles electron impact ionization of He$^{+}$ as a function of excess energy (energy above the ionization threshold of He$^{+}$) \cite{electronipmact}. While in the field-free case the impacting electron can have any initial energy, for the driven system the maximum return energy
to the core is 3.2 U$_{p}$ (three-step model), defining a different energy scale for the two processes. 

Given the above, for high total final energy, in the SI pathway, the impacting
electron does not transfer enough energy at the time of re-collision to the second electron to ionize both electrons. However, the presence of the field makes possible a transfer of energy to the second electron even after re-collision through the RESI pathway (Delayed pathway) thus casting the RESI pathway the prevailing one for high total energy.


{\it Asymptotic Observables.} The double differential probability of the energy sharing $|\epsilon_{1}-\epsilon_{2}|/(\epsilon_{1}+\epsilon_{2})$ versus the total electron energy reveals additional information for the DI process. For all four intensities currently considered we find that for increasing total energy the two electrons share the energy more asymmetrically compared to smaller energies. An asymmetry in energy sharing is also observed in electron impact ionization for high excess energy---many times referred to as U-shape energy sharing \cite{Ushape}.
 When the Delayed pathway prevails at high total energies, the energy sharing peaks around approximately 0.4-0.6, see \fig{fig:percent} third row. A study of the double energy differentials in energy (not shown here) reveals that the initially bound electron escapes with an energy around 2 $U_{p}$ (the energy a ``free" electron gains in the laser field). The re-colliding electron escapes with almost 3 times larger energy than the initially bound electron---the asymmetry is  larger for higher intensities due to stronger backscattering of the first electron resulting in its faster escape. Thus, this double differential probability can serve as a good predictor of the value of the total electron energy (for high energy) where the Delayed mechanism prevails. 
  
{\it RESIa versus RESIb.} In Table I, we show a break-up of the \% contribution of the DI mechanisms independent of energy. We also show results for 8.12 $\times$10$^{13}$ Watts/cm$^2$ at 1600 nm corresponding to the same Keldysh parameter as that for 3.25/13 $\times$10$^{14}$ Watts/cm$^2$ at 800/400 nm. Table I shows that when the Keldysh parameter is kept constant, smaller frequencies favor the RESIa versus the RESIb meaning that the initially bound electron ionizes mostly around a maximum and not a zero of the field. This is to be expected since the larger the frequency the faster the velocity changes around a zero of the field resulting in higher contribution of the RESIb pathway.

 \begin{table}
\begin{center}
\caption{\label{tab:table1}\% of DI mechanisms  over all DI events.\\
For each intensity, all mechanisms listed sum up to $\approx$ 100\%}


\begin{ruledtabular}


\begin{tabular}{c|cccc}

\multicolumn{5}{c} { }\\

\multicolumn{5}{c}{}\\


 8.12 $\times 10^{13}$ W/cm$^2$ at 1600 nm               & SI &  RESIa &RESIb &DE \\

\hline

 & 54.7&  27.2&16.5&0.8\\
\hline

\multicolumn{5}{c} { }\\

\multicolumn{5}{c}{}\\


 2.25 $\times 10^{14}$ W/cm$^2$ at 800 nm               & SI &  RESIa &RESIb &DE \\

\hline

 & 26.7 &  24.95&40.19&7\\

\hline


\multicolumn{5}{c} { }\\

\multicolumn{5}{c}{}\\


 3.25 $\times 10^{14}$ W/cm$^2$ at 800 nm               & SI &  RESIa &RESIb &DE \\

\hline

 & 60.8& 18.6&18.1&1.8\\
 
 \hline

\multicolumn{5}{c} { }\\

\multicolumn{5}{c}{}\\


 9 $\times 10^{14}$ W/cm$^2$ at 400 nm               & SI &  RESIa &RESIb &DE \\

\hline

 & 45.2 &  13.1&35.3&5.5\\

\hline


\multicolumn{5}{c} { }\\

\multicolumn{5}{c}{}\\


 13 $\times 10^{14}$ W/cm$^2$ at 400 nm               & SI &  RESIa &RESIb &DE \\

\hline

 & 53.6 &  15.41&26.5&3.53\\

\end{tabular}

\end{ruledtabular}

\end{center}
\end{table}

\subsection{Number of re-collisions and transfer of energy} 

We have found that for small total energy---where the TC pathway prevails---the transfer of energy from the re-colliding to the bound electron takes place first through a re-collision and subsequently through a triple collision. For the Direct and the Delayed pathway which have overall (independent of total energy) the largest contribution to DI the transfer of energy takes place through re-collisions. 

\begin{figure}[h]
 {\includegraphics[scale=0.27,clip=true]{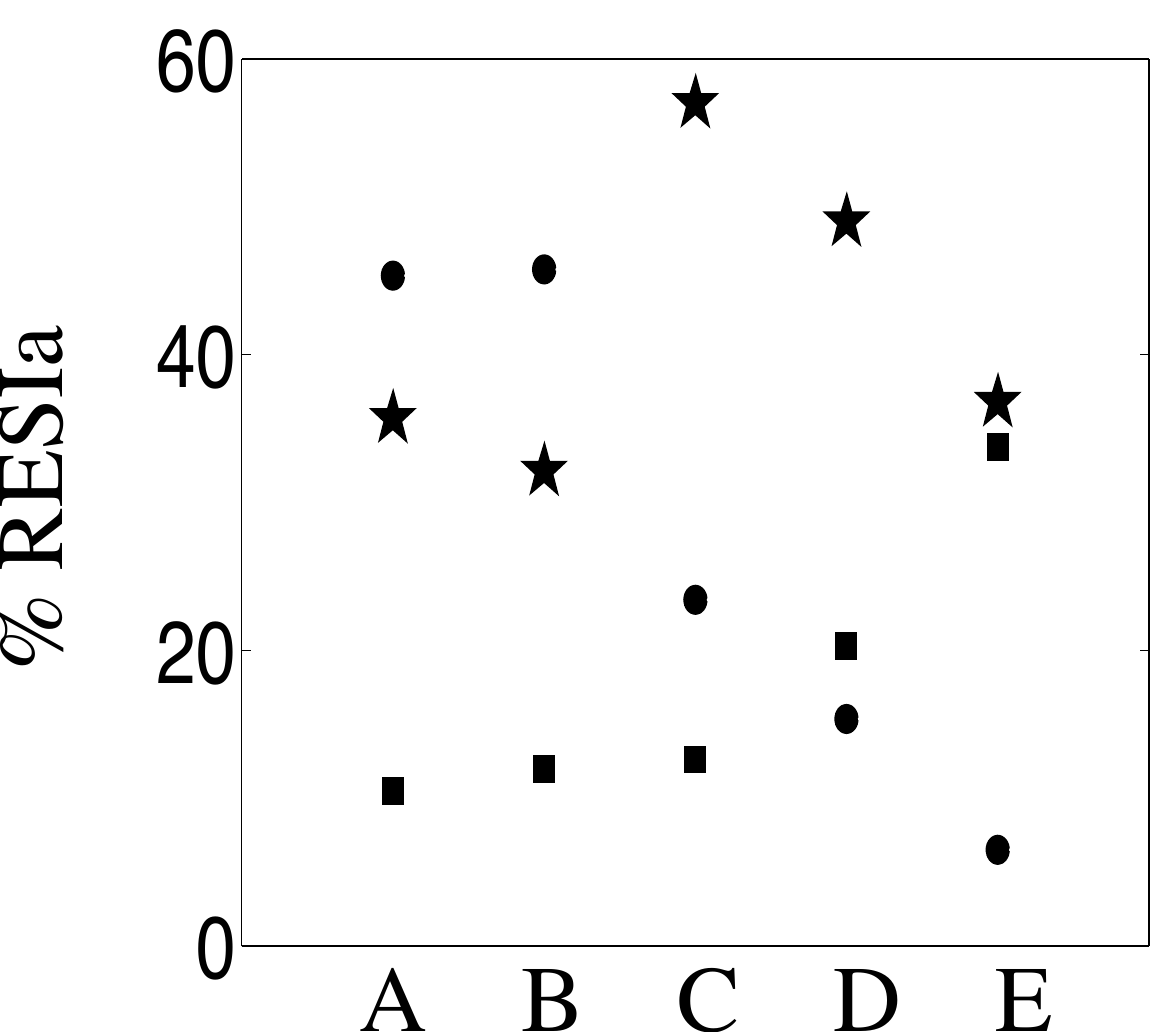}}
\caption{\label{fig:contributionRESIa}  The \% contribution of different number of re-collisions for the RESI mechanism out of all DI events. $\bullet$ for 1 re-collision,  $\star$ for 2 re-collisions, $\blacksquare$ for 3 re-collisions; Point A/B corresponds to 9/13$\times$10$^{14}$Watts/cm$^2$ at 400 nm, C/D to 
2.25/3.25$\times$10$^{14}$Watts/cm$^2$ at 800 nm, and E to 8.12$\times$10$^{13}$Watts/cm$^2$ at 1600 nm.}
\end{figure}

A legitimate question is how many re-collisions are required before the first electron transfers sufficient energy to the second one so that they both ionize and whether this number changes with frequency. \fig{fig:contributionRESIa} shows the number of re-collisions required for DI for the RESI pathway 
but similar results hold for the SI pathway as well. We find that at 400 nm the two electrons doubly ionize mostly after one re-collision with the second re-collision also contributing significantly to DI. The second re-collision has the largest contribution to DI at 800 nm.  Further decreasing the frequency to 1600 nm, we find 
   that the first re-collision's contribution to DI is only minor while the contribution of the third one is as prominent as the second's. The decreasing number of re-collisions with increasing frequency suggests
a more efficient transfer of energy to the initially bound electron for larger frequencies. Indeed,  the DI probability increases with increasing frequency: 2.4$\times$$10^{-4}$ for 8.12$\times$10$^{13}$Watts/cm$^2$ at 1600 nm, 3/8$\times$$10^{-4}$ for 2.25/3.25$\times$10$^{14}$Watts/cm$^2$ at 800 nm, and
2/3$\times$$10^{-3}$ for 9/13$\times$10$^{14}$Watts/cm$^2$ at 400 nm.  

\begin{figure}[h]
 {\includegraphics[scale=0.25,clip=true]{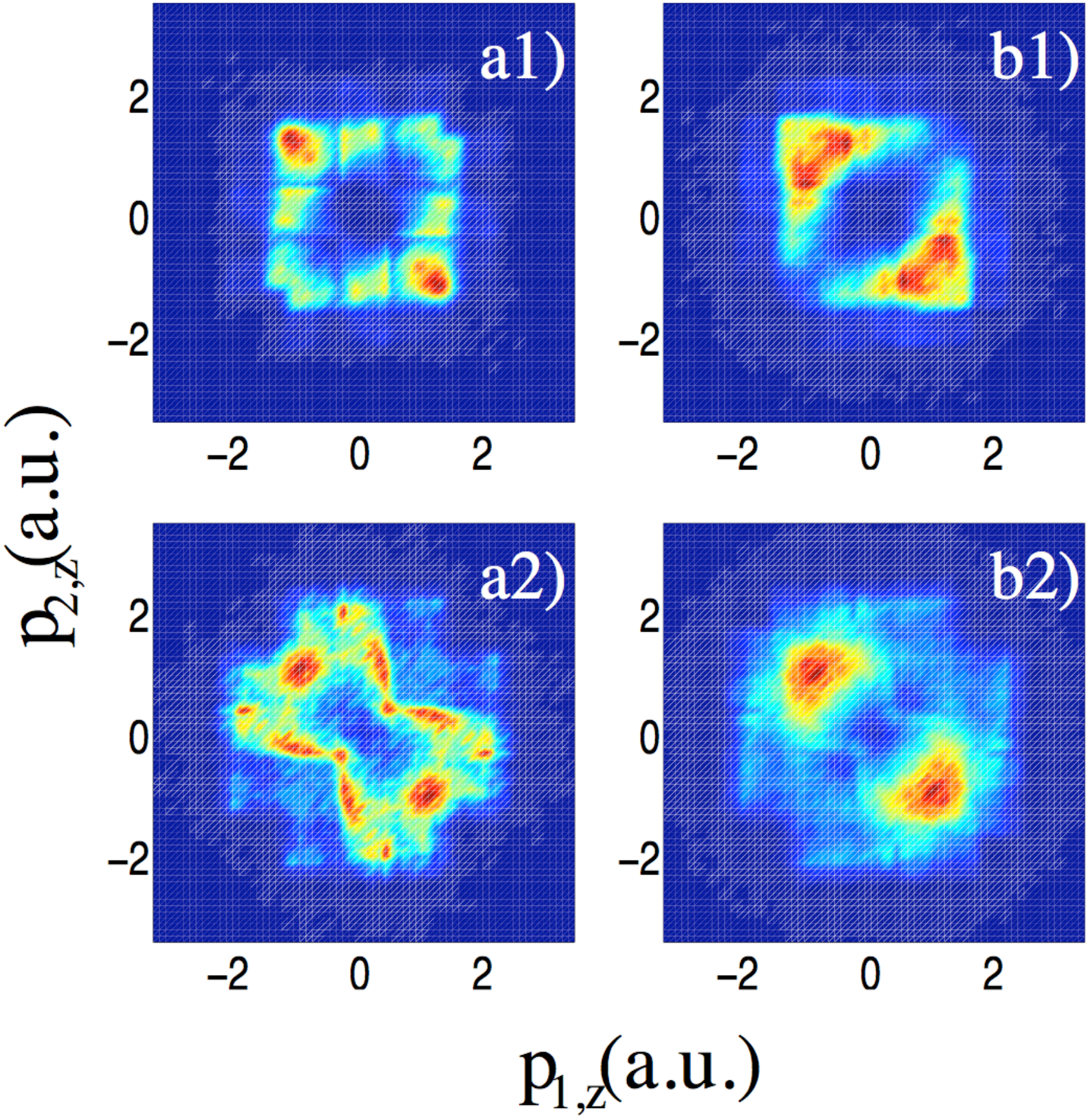}}
\caption{\label{fig:pz1pz2}  The correlated momenta for the RESIa mechanism (a) and RESIb mechanism (b); 1)  2.25$\times$10$^{14}$ Watts/cm$^2$ at 800 nm , 2) 9$\times$10$^{14}$ Watts/cm$^2$ at 400 nm.}
\end{figure}

 We find that this difference in the number of re-collisions affects the correlated momenta of the RESI pathway. Specifically,
 we see from \fig{fig:pz1pz2} that DI through the RESIa pathway at 400 nm favors cross-shaped correlated momenta along the two axes with the two electrons sharing the energy asymmetrically. At 800nm, DI through the RESIa pathway favors a structure in all four quadrants---as for 400 nm---however, the two electrons share the energy more symmetrically. A break down of the RESIa correlated momenta to contributions from different number of returns to the core of the re-colliding electron reveals that   
 DI after one re-collision favors the cross-shaped structure along the two axes. DI following more than one re-collisions favors a superposition of the two structures discussed above. Thus the prevalence of different number of re-collisions, before DI takes place, for different frequencies roughly explains why our RESIa correlated momenta are different for 400 nm and 800 nm.

\section{Conclusions}
We have shown that the prevalence of the Direct and Delayed pathway for intermediate and high total energy, respectively,  is common for laser pulses at 400 nm and 800 nm for a range of intensities. We have also identified TC as a new DI pathway that prevails for small total energy. Three-body collisions between the nucleus and the two electrons underly the TC pathway. Given the small energy available to the system the two electrons escape in opposite directions giving rise to distinct patterns of the correlated sum of the momenta components.

 \vspace{2cm}
{\bf Acknowledgments.}
We gratefully acknowledge Thomas Pattard for a critical reading of the manuscript and helpful discussions as well as funding from EPSRC (grant no. EPSRC/H0031771) and NSF
(grant no. NSF/0855403).
 

\bibliographystyle{unsrt}

\end{document}